\documentclass[a4paper,11pt]{article}
\pdfoutput=1 

\usepackage{jheppub} 

\usepackage[T1]{fontenc} 

\usepackage{subcaption}

\title{On the possible implications of Dark Matter in the rings of Saturn: a conjecture}
\dedicated{In memory of Professor Horia Hulubei.\\ \textbf{This paper is an extended version of a presentation \\at a webinar organised by the Romanian Academy \\ November 15, 2021.}}

\author[a]{Alexandre Ciulli}
\author[b,1]{and Sorin Ciulli \note{Corresponding author.}}

\affiliation[a]{University of Clermont-Auvergne,  France}
\affiliation[b]{University of  Montpellier,  France}

\emailAdd{alexandre.ciulli@hotmail.fr}
\emailAdd{ciullisorin@yahoo.fr}

\abstract{In this article we discuss some consequences of the well-known proposition of Fritz Zwicky  ~\cite{zwicky1}, published in the nineteen thirties, that Dark Matter `mimics' the inertia-gravitational behaviour of usual matter.  In particular, we consider some special dynamical regions such as those of  the  Ring Systems of the gaseous giants at the edge of the Planetary System. This article is a continuation of an earlier paper  ~\cite{sebuciulli},   where it was shown that gravitationally interacting particles may remain near the Lagrange Points L4 and L5 for many thousands of years. This provides enough time for the Dark Matter, if present there, to interact with the usual matter.  We discuss also a number of questions related to places which might be considered  singular in the mathematical sense. } 
\keywords{Dark Matter inside Solar System,  Rings of Saturn.}

\begin{document} 
\maketitle
\flushbottom

\section{Introduction. Are the visible Rings of the Giant Planets accompanied by some invisible ones?}
\label{sec:intro}
\begin{figure}[tbp]
\centering 
\includegraphics[scale=0.4]{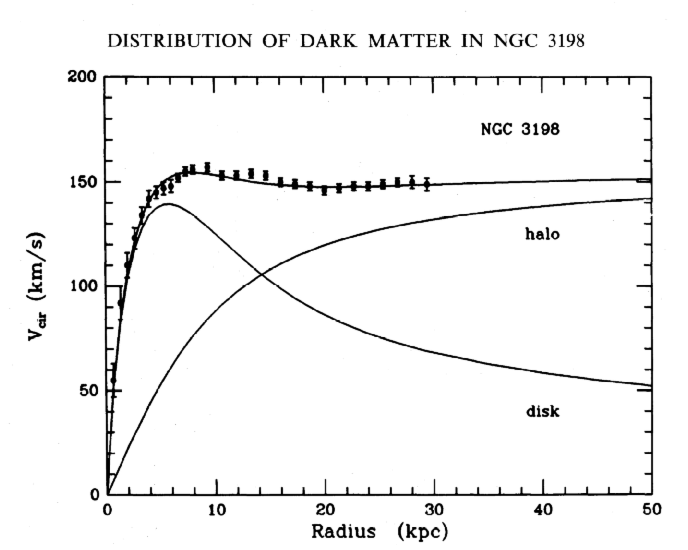}
\caption{\label{fig_galaxy_rotation_DM} Rotation of stars from the Spiral Galaxy NCG 31 98 (Van Albada 	et al. 1985, see ~\cite{abdalaDM1985}).
The upper curve with error bars shows the experimental  data for the velocities  (from Doppler shifts) of the individual stars, while that labelled 'disk' represents the Kepler motion one might calculate from the amount of visible matter present  in the disk of the  galaxy. To recover the observed velocities one has to add an invisible halo containing much more material (Dark Matter) than the visible one. Distances are in kilo-parsecs, velocities in km/sec.}
\end{figure}

Recent findings of the NASA spaceship Cassini  ~\cite{cassini_expedition} have shown that the Rings of Saturn are extremely thin, something like 10 metres across, in comparison to the dimensions of the planet itself and the other nearby astronomical objects. Now according to the daring proposition made by Fritz Zwicky   ~\cite{zwicky1}  (for many years at the Palomar Observatory, USA)  and also of his friend and collaborator, Walter Baade (from the Mount Wilson Observatory), usual matter should be accompanied by a much larger amount of an elusive and invisible material, which  should have inertial and gravitational properties similar to usual matter. Further, Jan Henrik Oort \cite{Oort}, Ernst Öpik ~\cite{Opik1915}, Vera Rubin ~\cite{Rubin1983}) have suggested something similar in order that the galaxies should be stable and do not blow up as a result of the centrifugal forces acting on them  -- see figure  \ref{fig_galaxy_rotation_DM} or/and Reference ~\cite{abdalaDM1985}. To explain why this material has not already been seen experimentally, Zwicky and Baade made the simplest possible hypothesis, namely that although this substance might be gravitationally as active as usual matter, electromagnetically it should be absolutely transparent to normal light. So the concept of Dark Matter appeared. Indeed when Astronomers look at the night sky and see just nothing, the sky for them  is Dark! 

        

Coming now  back  to the (visible) Rings of Saturn or to those of the other Gaseous Giants at the edge of our Solar System, according to this conjecture they might  be accompanied by some other invisible Rings that share similar, if not identical, positions. Indeed all four planets, Jupiter, Saturn, Uranus and Neptune have some sharp rings around them. Now we should  not repeat here the over-simplification which Boltzmann made in the 19th Century in a different context and assume that Dark Matter shares  exactly the same position as the usual matter, but might be present in many places in our Solar System. It has been argued that all the Dark Matter around us has been already expelled from our Solar System in the past by some kind of sling effect, like the dust and the other small objects swarming in the sky. We do not believe that this is likely to be so, especially in the distant regions of our Solar System where these sling effects which are related to the centrifugal forces are much reduced. Indeed, the very existence of the dust as well as of the other small constituents like pieces of ice and pebbles which can be seen in the images taken by the rocket Cassini ~\cite{cassini_expedition} in the Rings region, prove that this is not really so. However if there is not enough Dark Matter  in Saturn's Rings region, we can move our whole discussion further, towards the Kuiper Belt. Uranus and particularly Neptune have their own Ring systems which again seem to be extremely sharp. The Kepler velocity around Neptune is approximately 5.4 km/s which is roughly half of the  9.7 km/sec for Saturn (see, for instance, NASA's planetary data ~\cite{nasa_planetary_data}). Since sling effects depend on the centrifugal forces, they are proportional to the square of the velocities (see further the example discussed in the first footnote below), and so the corresponding  sling effects will be considerably weaker there.

 A direct argument that Dark Matter cannot have been expelled completely by simple sling effects from a region as large as our Planetary System, comes from the existence of the visible Rings themselves, since the visible Rings (NASA, see Cassini \cite{cassini_expedition}) contain small lumps and ice pellets (and powders) which clearly have not been expelled at all. So one could ask why it is that only the Dark Matter has been expelled!

\section{Where Dark Matter might be located}

\begin{figure}[tbp]
\centering 
\includegraphics[scale=0.6]{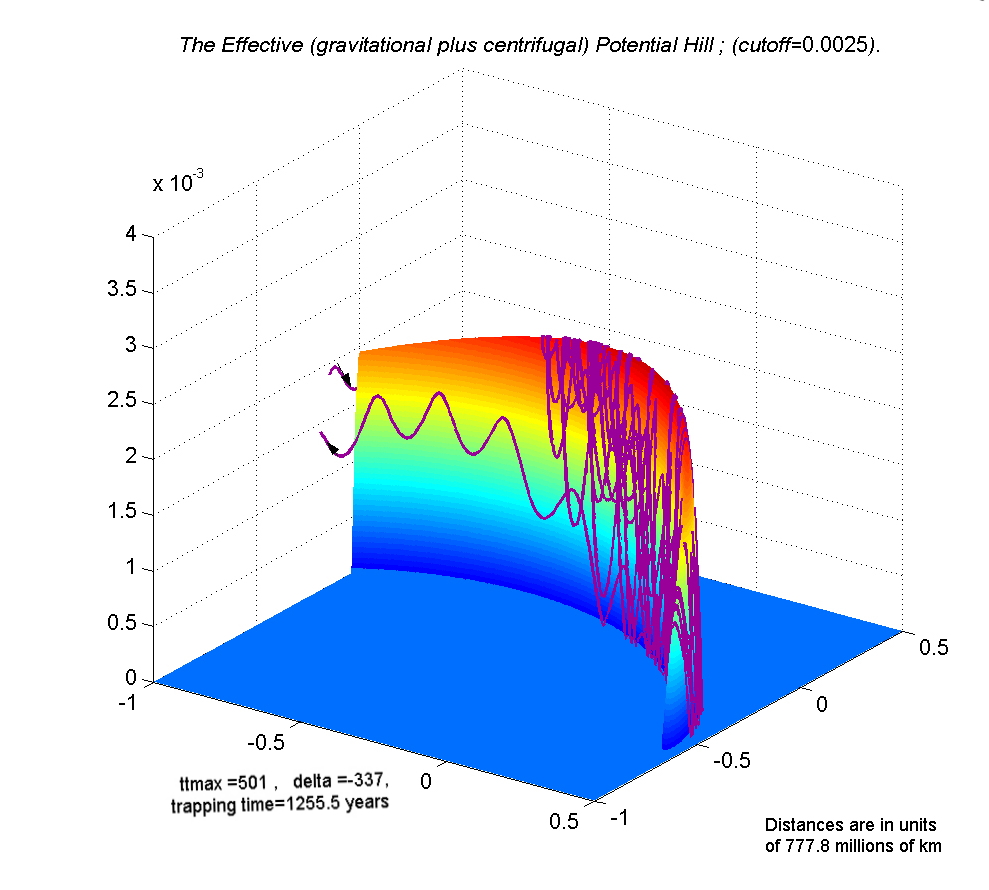}
\caption{\label{fig_lagrange} A computed Dark Matter (or usual Matter) particle trajectory  around the  Lagrange Point L4 of Jupiter \cite{sebuciulli}. The origins  of these symmetry breaking produces the accumulations of loops and knots (extremal points) on the trajectory, which last thousands of years, cannot be attributed to trivial energetic causes, since the trajectory passes above the maximum of the pseudo-potential. They are dynamical consequences of the Coriolis force around L4 and L5.}
\end{figure}

According to what we know about Dark Matter, apart from gravitational effects  its intrinsic interaction with  usual matter is extremely weak.  However, by Zwicky's conjecture,  Dark  Matter should have similar inertia-gravitational properties to  the usual one and so it may accumulate in  various  zones of our Solar System, such as in the neighbourhood of some Lagrange points like L4 and L5, where it is known that usual matter also accumulates. See, for example reference \cite{sebuciulli} where it was shown that the trajectories of usual matter may spend thousands of years in the corresponding Lagrange points of Jupiter (see below figure \ref{fig_lagrange}). Hence, according to the ansatz of Zwicky, the dynamical forces that lead to the existence of the visible Rings of Saturn should act in a similar way on Dark Matter, and hence the visible Rings should be accompanied by some totally invisible ones. Hence, Dark Matter could share (almost) identical positions with the visible matter, unless some so far unknown Strong Exclusion Principles forbids this completely. The observational fact that the visible Rings are so thin in comparison with the other cosmological objects that we are used to -- 10 metres in some places -- suggests that if some Dark Matter is also present there, it should be closely mixed with the usual matter.

\begin{figure}[tbp]
\centering
\includegraphics[scale=0.2]{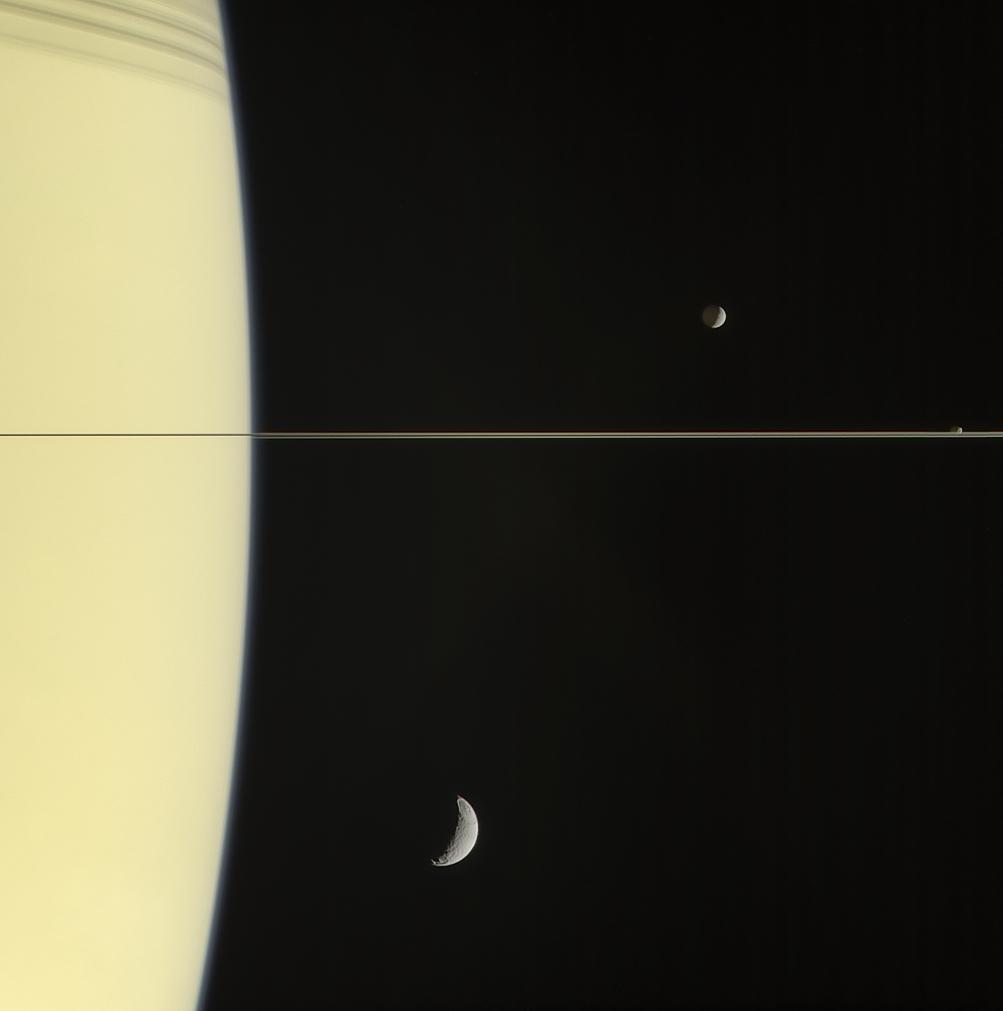}
\caption{Edge-On View of Saturn's Rings: Cassini's narrow-angle camera, 2006 (courtesy NASA/JPL-Caltech).  Owing to their extreme thinness we  may be tempted to call them ''singular'', but since they  contain observational inacurracies we shall use the expression ''quasi singular''. The reader should compare their size with that of the Planet.
 \label{fig_cassini_saturn_edge_on}}
\end{figure}

There exists also this half not-expressed belief, that in our world  new qualities may arise in the neighbourhood of singularities and it as is well known, René Thom \cite{rene_thom} has widely advocated these ideas. This is mainly because singularities are usually restricted to small regions of space and hence the various elements have an enhanced opportunity to mix thoroughly.

But who decides when and where singularities occur ?  In Physics when you meet an object only a few metres across, while around it are others with dimensions of thousands of kilometres, you  may be tempted to call it singular. Anyhow if some Dark Matter were present there, owing to the thinness of the Rings, it would be well mixed with the normal matter. We are in fact in the usual conditions of a Nucleophilic Substitution Reactions S$_{N}$2, between two substances, where the overall yield  is proportional to the product of the densities of the reactants. Owing to the lack of any reliable information on the nuclear interaction lengths between the two kinds of matter, as a first approximation we suppose that the output is proportional to the product of the concentrations.

So far there are no reasonable data concerning the cross sections between these two kinds of matter but, owing to the fact that the Rings are extremely thin, and that their extension is extremely large, by waiting a sufficiently long time we might expect to find some interaction between these constituents. To give some figures on the quantities involved here, we should consider an annulus with an inner radius of approximately 62628 km which is the equatorial radius of Saturn, and an outer radius equal to that of the outermost Ring, estimated at 483000 km.  Since the distant Rings F and  E  have a much  lower density, in a first approximation they might be completely neglected, so that we may stop this evaluation much  earlier, say, at some 130000 km to consider only the dense Rings (i.e. the Rings  A and B, as well as those which are nearest to the  Planet, C and D). We will then have to subtract some 10\% or 20\%  from this area since there is a gap between the planet and the first Ring, D, as well as the surface area of the main gaps ENCKE and CASSINI produced by the principal shepherd satellites\note{As is well known, the (`classical') gaps visible at the surface of the Rings are consequences of the fact that if a small object moves on the same orbit as a shepherd satellite (a satellite  moving among the Rings) but in the opposite direction, it will be thrown away as the result of the gravitational interaction with this satellite. As an example consider a small object (a golf ball, for instance) with a velocity $V$ coming from the opposite direction along the same orbit as a given shepherd satellite. According to Newton's laws (in a first approximation we consider only circular orbits) the velocity of this small object will have the same modulus as that of the satellite coming  in the opposite direction, so that their relative velocity will be 2$V$.  If they are close enough, the small object will be attracted by the shepherd satellite, so that it will go around it, and then leave the scene with a velocity 2$V$ in their  centre of mass system. However, as  the satellite is much heavier than the golf ball, this centre of mass will almost coincide with that of the satellite alone. Since the golf ball now moves with a speed 2V relative to  the satellite, it has a velocity 3V with respect to Saturn.  As the centrifugal force is proportional to the square of the velocity, this force will be nine times larger than that corresponding to its  stable orbit. A factor as large as nine will be enough to expel the small object far away, perhaps as far as the Kuiper Belt.  
\label{footnote_shepherd}}. The particles produced in these reactions may then be observed by waiting a sufficiently long time using devices placed at the surface of some satellites  like Enceladus, where further NASA expeditions seem to be planned.
			
Owing to the lack of reliable information on the interaction lengths between the two kinds of matter, as  a first approximation we shall suppose that the yield of the reaction is proportional to the product of the concentration of the two substances. The particles produced in these reactions may then be observed by waiting sufficient long time, by devices placed at the surface of some co-planar satellites, like Enceladus.

\begin{figure}[tbp]
\centering
\includegraphics[width=14cm,height=8cm]{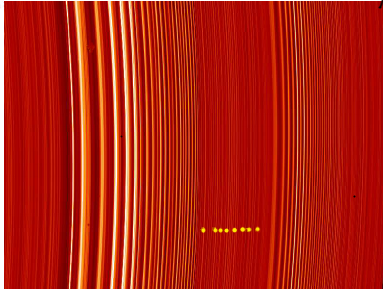}
\caption{A Cassini picture of the rings, in sunset light (courtesy NASA/JPL-Caltech). One can see the waves produced by the shepherd satellites, but also the extremely regular fine lines marked 	here with dots that are not yet explained in a satisfactory  way. \label{fig_cassini_saturnrings}
}
\end{figure}

\section{The ``Fence Effect''}

 In figure \ref{fig_cassini_saturnrings} above, we have deliberately chosen a region which is a little apart from the waves produced by the shepherd satellites. This is to avoid being disturbed by their presence and hence enable us to see better the ''crude'' interaction of their would be multiple ''spectra''.  See below for explanations. Specifically, we should observe (in figure \ref{fig_cassini_saturnrings}) the very fine lighter lines orthogonal to  the radial direction of the Rings,  i.e. along the direction of the Rings' coordinate system, marked with yellow dots.  These might be interpreted as a kind of ''Zaun Effect''  (in German,  or ``Fence Effect'' in English, for explanations see below footnote\note{To explain this in a simple way, we shall send the reader to Switzerland, a country in which, being  a bit wealthier, the intersections  between  roads of different ranks are always made by  means of bridges. The bridges usually have two fences, and the front balusters appearing to an observer approaching the bridge from beneath, a little larger because of  perspective, will occult at regular intervals the rear ones. So since in these places only one, and not two, baluster are visible they will appear as lighter fine lines at regular spacings. These lines are  also an  indirect indication that here we have to consider the existence not only of a single `fluid',  but of two or  more.

 This is the origin of the moir\'e patterns which do appear when we look to two sheets of fine silk put together, which obviously contain more information  than a simple optical interference.  Although the example given here has to be considered just as a  metonymy,  much of the geometric features of the ``balusters" can be  inferred from the geometrical widths of the fine occlusion lines,  or from the  (slow) variations of  their positions and spacings.   So quite a lot of information might be  obtained  by  a careful inspection of these images. \label{footnote_fence_effect}}),  between the "spectra" produced by  two  (or  more)  different objects (or fluids)  which might be simultaneously present there. These phenomena may look at a first glance similar to the usual interferences of two waves, but, as is  discussed in footnote \ref{footnote_fence_effect}), its nature extends far beyond the usual interferences. They may be related to different causes, as for instance to the so called  moir\'e patterns produced by two sheets of fine silk put together.

 The fine lighter vertical lines might represent a direct proof of the existence of  a second,  if  not of a multiple Ring system instead of a single one. Considering the regular patterns visible in figure  \ref{fig_cassini_saturnrings}, we might have  the impression that we are observing a carefully printed table from some mathematical treatise on Spherical, or, more correctly, Two Dimensional Cylindrical Functions. It is our belief, which we tried to express in this article, that such special Dynamical  Regions like that of the Rings  --  which are extremely sharp, ``quasi singular"  --  or that of the Lagrange Points discussed in Ref. \cite{sebuciulli} where the trajectories remain for long periods of time, are zones where these substances  mix and where future experiments should be planned.

Some people who try to find various elusive structures in between the fine vertical lines appearing in the figure \ref{fig_cassini_saturnrings}  (for instance among  the lines 1\&2, 3\&4, 4\&5 and 7\&8, marked by dots),  might  get the impression that they see the  patterns of  some Helmholtz-Kelvin instabilities between two intermixed fluids. A brief comment: Helmholtz-Kelvin or similar singularities seen in the flows of various fluids may be recognisable by their form but not by a ready to use analytic algorithm.  This situation  is somehow similar to that of the examination of  rocks ejected by the volcanoes from Mars (or Venus) which fall on Earth as meteorites after a long interplanetary trip. According  to our geological colleagues who organise yearly scientific trips to the Alto Plano ATACAMA  to collect them, there does not yet exist any algorithm to recognise them among the many other rocks.  Only a long human experience permits to accomplish this job. We are probably in the same situation when we try to recognise cosmic singularities looking only at their physical appearance.

\begin{figure}[tbp]
\centering
\includegraphics[scale=0.05]{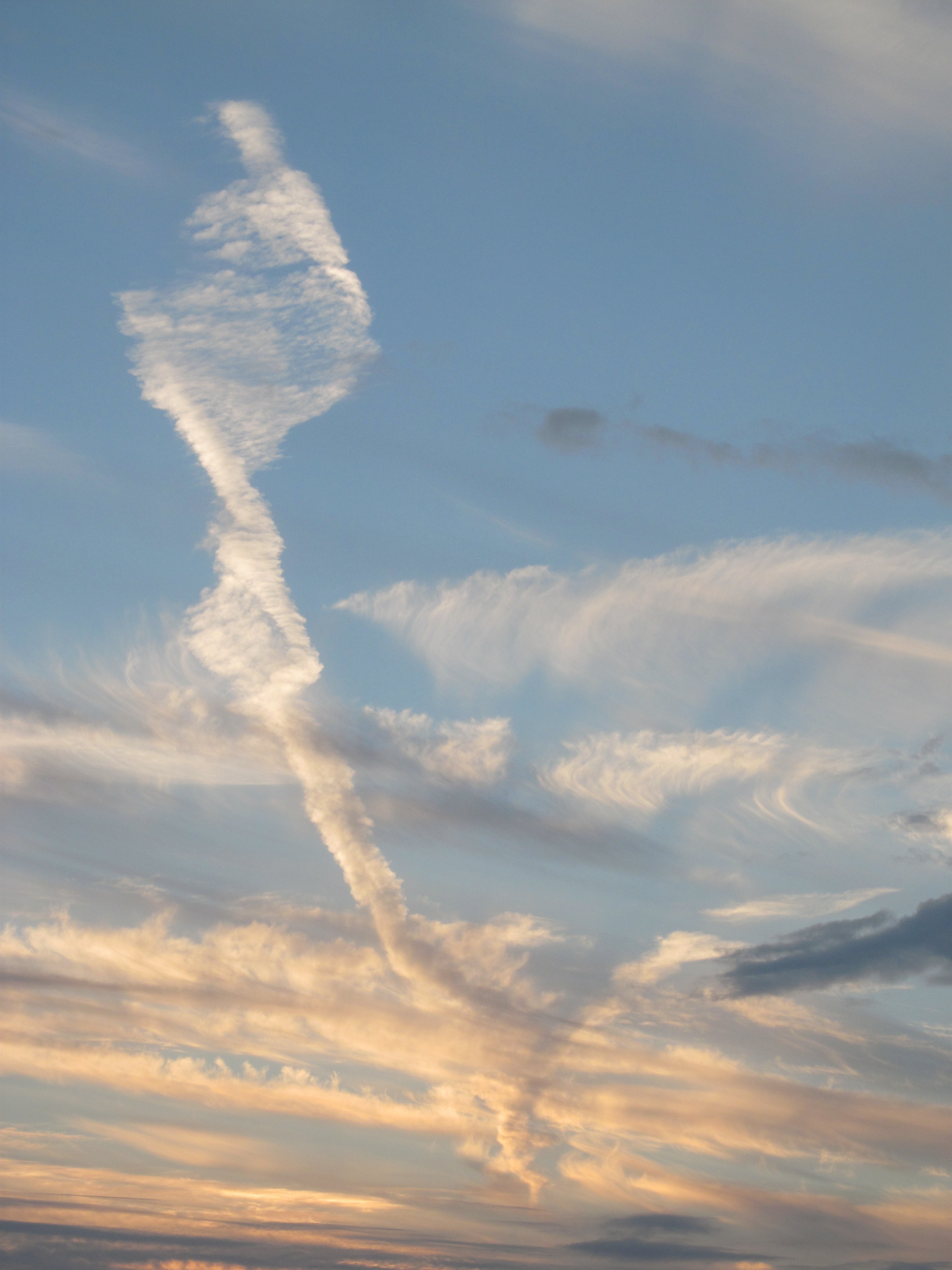}
\caption{Helmholz-Kelvin singularity.  If by chance you have once looked at the sky overhead and seen the clouds entangled as in this double helix, you have possibly seen a Helmholz-Kelvin singularity.
\label{fig_helmholtz_kelvin}
}
\end{figure}

\begin{figure}[tbp]
\centering
\includegraphics[scale=0.085]{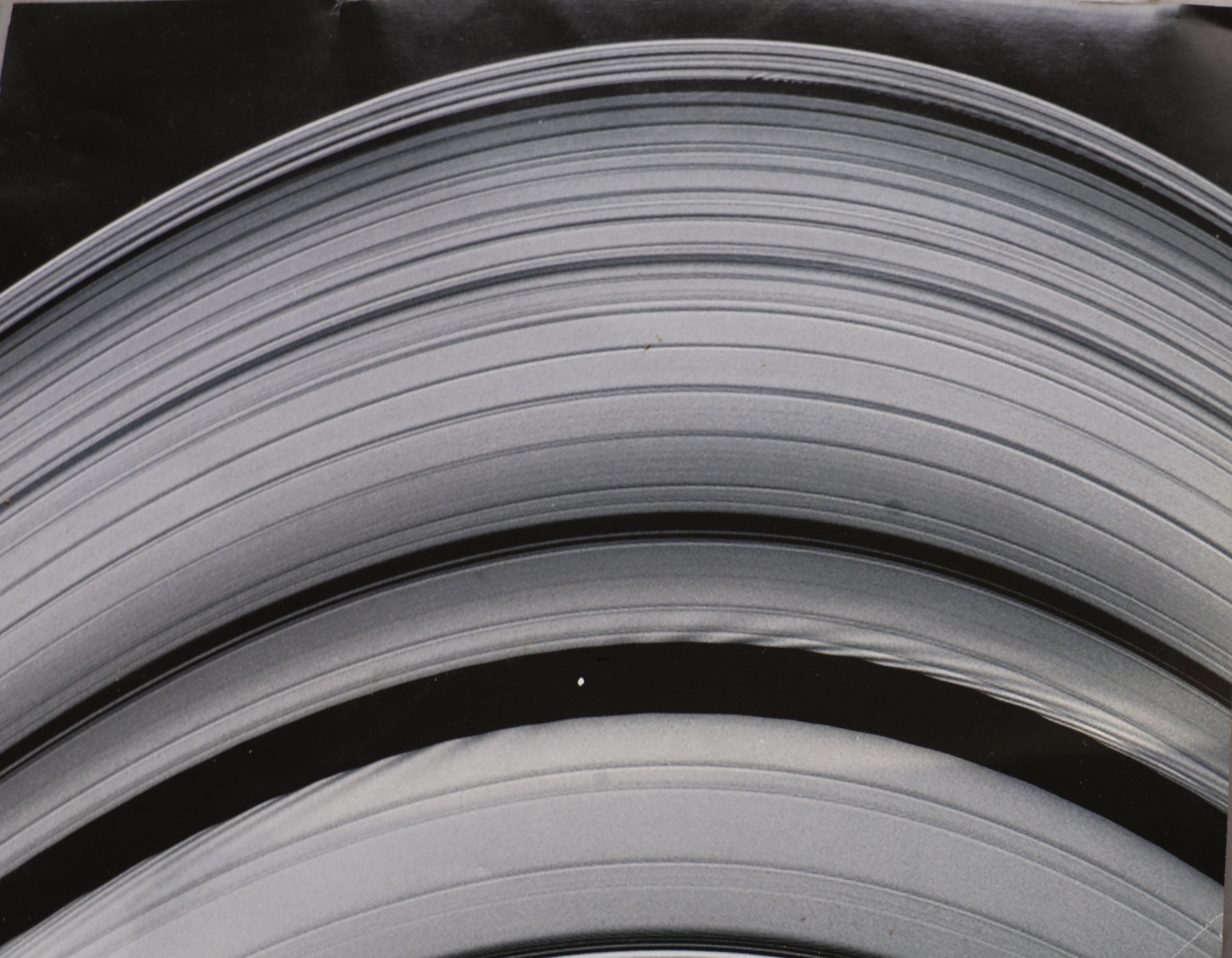}
\caption{NASA's  picture of the Rings, with the "ditches"  (and the somewhat larger "moats") as well as the ``classical, black"  (see main text) gaps (courtesy NASA/JPL-Caltech).  One can also see  the  shepherd satellite PAN in the middle of the gap ENCKE and also the  retarded/advanced  wavelets induced on the Rings, (``advanced",  since the Kepler velocity of the inner ring is larger that the  outer preceding one).
\label{fig_saturn_rings}
}
\end{figure}

Certainly there are sufficient unexplained features in these images to justify the examination of new ones. We have, for instance, not yet  mentioned the existence of the radial Spikes described in the NASA reports \cite{cassini_expedition}, which ``run" on the surface of the Rings. However, here we may be tempted to interpret them as moire effects (see footnote \ref{footnote_fence_effect}), as those discussed in the text.  This, at the same time, could be an explanation for their short life nature. Anyhow the rich information brought by the Cassini  mission will certainly support research activities for long periods of time.

\section{A Powder Dynamics approach.}

There are also other facts that might have some importance in the understanding of  the formation and the history (the ``geology") of the Rings. There are (a) the multiple parallel  stripes or circular annuli, very visible in another excellent NASA picture, figure \ref{fig_saturn_rings}. They are also present as the fine lines, discussed in figure \ref{fig_cassini_saturnrings} in the previous section, and, very importantly, they have no transparent background (they are not ``black", in contrast to the classical gaps through which you can see the surrounding, dark Space). Then (b), there are the classical gaps produced, see Footnote \ref{footnote_shepherd}, by sling effect by the shepherd satellites, which have transparent backgrounds and hence they appear black in the pictures.

As the reader may notice in figure \ref{fig_saturn_rings}, the widths of the circular annuli vary, monotonically increasing or decreasing with their positions. A theoretical explanation based on Powder Dynamics ~\cite{Rietema} of such effect will be the subject of a subsequent paper.  This remark represents an important tool which shows that the ``classical" (the ``black") gaps were formed at a later date than the circular annuli, since as the reader may verify, they do not interfere at all with the monotonicity of the widths. This is an indication that when the black gap appeared, the ``ditches" and ``moats" were certainly already there. 

\begin{figure}[tbp]
\centering
\includegraphics[scale=0.5]{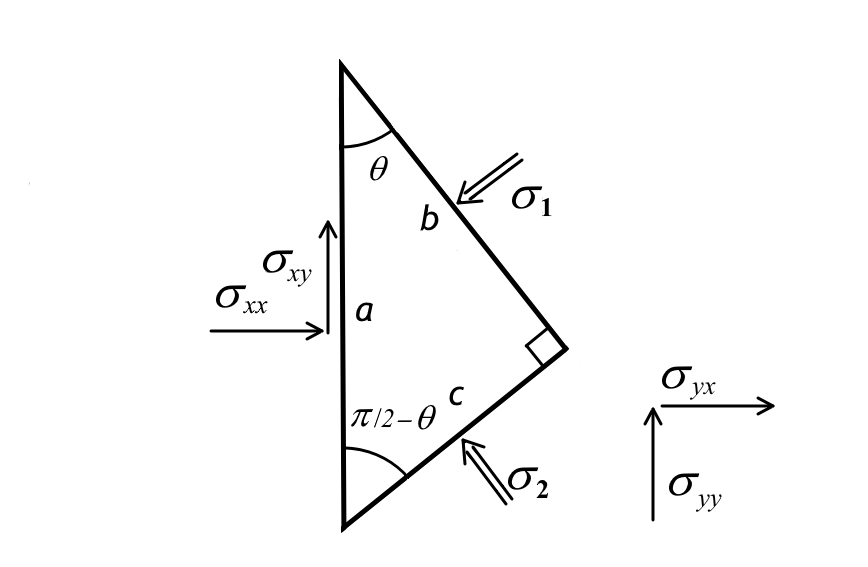}
\caption{Mohr's Triangle. Schematic picture of the stresses acting on an infinitesimal triangle at the surface of the Rings. The initial eigenvectors stresses depict with double arrows and are mutually orthogonal, as they should be. The $xx$ and $yy$ components represent the normal stresses  while the $xy$ and $yx$ ones  the shear ones. We  purposely tried as much as possible to stick at the historical notations (i.e. to the ``circle of Otto Mohr").
\label{fig_mohr}
}
\end{figure}

Since the Rings of Saturn are extremely thin we find ourselves in  a two dimensional situation. Care should however be exercised to introduce the initial conditions correctly. While in three dimensions the forces are obtained by products of the area with the pressure acting on it, in two dimensions we will have to multiply lengths of segments with the corresponding components of the stress tensor.   We shall then follow the theory of Mechanics of Powders at Rest for a two-dimensional system presented in ~\cite{Rietema}. To this end, we start by considering a force acting perpendicular to the side of  length $a$ of an infinitesimal triangle at some given point P on the surface of the Rings (see figure \ref{fig_mohr}).  This
force is related to the  $\sigma _{{xx}}$ component of the stress tensor (we have striven ourselves  as much as possible to follow the notations of Mohr). The first of  these two subscripts indicates that it is parallel to the $x$-axis and that it acts on the first segment of length a of the triangle from figure \ref{fig_mohr}. If the second index were $y$, this would  mean that we are considering the orthogonal component of the force i. e. that which is parallel to this segment  $a$.  Since the stress tensor is Hermitian  (the stress matrix being  real and symmetrical) the shear components  of the stress  acting in the orthogonal direction, have to be equal $\sigma _{{yx}}=\sigma
_{{xy}}$.

Let $\sigma_1$  and  $\sigma_2$ denote the \emph{initial eigenvectors} of the stress tensor acting on this infinitesimal triangle, namely perpendicular on the segments of length $b$ and $c$, producing
respectively the forces which act on these sides (represented with double arrows in figure \ref{fig_mohr}). Projecting now
everything onto the x axis, the equation of the balance of the forces along the $x$ direction is :

\begin{equation}
    a \cdot \sigma _{xx}=b \cdot \sigma_1\cos(\theta )+c \cdot \sigma_2\sin (\theta ).
\end{equation}

 Both  $b$ and $c$ can themselves be expressed in terms of $a$ and the angle $\theta$ : 
 
 \begin{equation}
     b=a\cos (\theta )  \mathrm{\ \ \ and \ \ \ } c=a\sin (\theta),
     \label{eqn_trig2}
 \end{equation}
 that after simplification we obtain the following equation \note{All that can also be represented very nicely in a graphical form by means of the circle of  Christian Otto Mohr, a  nineteenth century German engineer,  Professor at the  Universities of Stuttgart and Dresden,  much beloved by his students, who received for his work the highest distinction from the Fuerst of the State of Sachse. \label{footnote_mohr}}:
 
 \begin{equation}
     \sigma_{xx}=\sigma_1 \cdot \cos ^2(\theta )+\sigma _2 \cdot \sin ^2(\theta).
     \label{eqn_sigma_xx}
 \end{equation}
 
 Now, using basic trigonometric identities, we obtain :
 
 \begin{equation}
     \sigma_{xx}=\frac{1}{2}(\sigma_1+\sigma _2)+\frac{1}{2}(\sigma _1-\sigma _2)\cos (2\theta).
 \end{equation}
 
Considering now the shear component  $\sigma_{xy}$ of the stress tensor which acts along the segment orthogonal to the $x$ axis, from figure \ref{fig_mohr} we obtain :

\begin{equation}
    a \cdot \sigma _{xy}=b \cdot \sigma_1 \cdot \sin (\theta ) - c \cdot \sigma_2 \cdot \cos (\theta ).
\end{equation}

The minus sign appears here since the $y$-components of the two initial eigenvectors stresses point in opposite directions.  From \eqref{eqn_trig2} and simplifying with $a$, we have

\begin{equation*}
\sigma _{{xy}}=(\sigma _1-\sigma _2) \cdot \cos (\theta ) \cdot \sin (\theta ),
\end{equation*}
which means that

\begin{equation}
    \sigma_{{xy}}=\frac{1}{2}(\sigma _1-\sigma _2)\cdot \sin (2\theta ).
\end{equation}

Similarly,  $\sigma _{{yy}}$ can be computed in an analogous way. Or much more quickly, using the invariance properties of the \emph{Trace} :

\begin{equation}
    \mathrm{Tr}(\sigma) =\sigma _1+\sigma _2=\sigma _{xx}+\sigma_{yy}
\end{equation}

\begin{equation}
    \sigma _{yy}=\frac{1}{2}(\sigma_1+\sigma _2)-\frac{1}{2}(\sigma _1-\sigma _2)\cdot \cos (2\theta)
\end{equation}

In a paper in preparation, we shall use these equations to analyse the monotonicity of the widths of the circular strips visible at the surface of the Rings. Before that we should probably say also few words about the microscopic forces acting among the constituent particles of the Rings. This task is relatively simple since the pictures provided by Cassini show clearly that the distances among the particles are rather large.  In general the form of the attractive potential derived from the induced dipoles due to the  London dispersion forces active there, have a very short range of order $-\mathit{const} \cdot \frac{1}{r^6}$. The Cassini pictures show clearly that the density of the powders is not very  large, the distances  between the grains exceeding by much these ranges.  There exists of course also a  repulsive component, but this has a much shorter range,  the  usual form given for the Lennard-Jones potential being  $(r_0/r)^{12}-(r_0/r)^6$. The representation of this  repulsive part is not considered to be realistic, an exponential form $e^\frac{-r}{r_0})$ being  much  closer to the decay of the wave function and hence  preferred for the repulsive part. We shall address  these questions in our forthcoming paper. And who knows, the prevalent attractive potential acting between the ``boulders" (ice pellets, grains of sand)  above a certain size existing in the Rings, might be nothing else  than the usual Newtonian potential, i.e  the  $\frac{1}{r^2}$ force.

\section{Some concluding remarks}
In summary,  according to Zwicky's classical conjecture Dark Matter mimics the gravitational and the inertial behaviour of usual matter. So it is always interesting to find gaps in the Rings region, even very narrow ones and to look carefully at its borders to see whether the advanced/retarded  wavelet system is present and to verify whether the shepherd satellite is visible or not. If the gap under consideration is a ``classical black gap" -- see the previous section --  usually it should be present!

To our knowledge these features seem to  have  been  somehow overlooked, at least in previous studies, since until now Dark Matter has been studied more at the Cosmic Level. To be fair, however, some of these findings, for instance those discussed in figure \ref{fig_bullet_cluster} below, have the merit to have much enhanced our confidence in the existence of Dark Matter. In the future we shall probably be back in these (almost) singular regions of our Solar System, with suitable devices to be placed on Enceladus or on some other shepherd satellites to make measurements.

The discussions in this paper are based tacitly on the belief of Ren\'e Thom \cite{rene_thom}  that  `Singular'  Dynamical Regions may be able to produce such unusual (very thin) objects like the Rings of Saturn. They are really extremely `thin'  in comparison with all the cosmic objects around, so they might represent preferred places where the two kinds of matter could mix, and  hence, interact.  Another important approach is, of course, that of Sergio Bertolucci et al. \cite{bertolucci} (see also references therein) in which  one looks at places where also usual matter accumulates naturally, like the interior of the Sun or the large astronomical objects. Indeed according to the conjuncture of Zwicky, Dark Matter  follows the patterns of usual one and so  these regions will be in the first row where one could  find  interactions. 

 Before closing this section, we should like to add also some words which, although lying outside of the  main topics of this paper, are interesting by themselves. This is particularly true as NASA apparently plans to return with a new expedition on Enceladus, with a small submarine containing a nuclear source of energy to melt a hole through the glacier crust and to explore the liquid sea beneath it  in order to look for possible  forms of Life. Indeed, while the northern hemisphere of the small satellite Enceladus has the usual marks (craters) of various impacts, the southern one is absolutely smooth, as after a recent snowfall. Further South, near the South Pole of this moon, Cassini's cameras have detected the plumes of some Giant Geysers, rising high vertically, since the gravitational field of these moons is rather weak. See figure \ref{fig_geyser_enceladus}  below. This is direct proof that under the glacial crust we find liquid water at least around zero degrees Celsius.  As Saturn is rather far from the Sun, it is clear that this heat is not provided by the Sun, but probably is produced  by a mechanism based on the tidal forces made by Saturn which is not too far away.


\begin{figure}[tbp]
\centering
\includegraphics[scale=0.35]{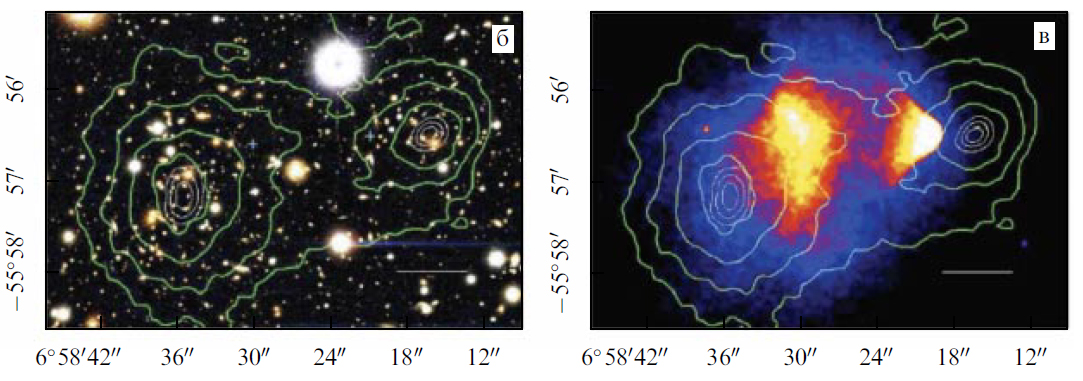}
\caption{Bullet Cluster : visible image, UV and gravitational lensing observations (courtesy NASA/JPL-Caltech). Probably one of the most striking events related to Dark Matter, at the Cosmic level, is	the `Bullet Cluster Galaxies Event', recorded by the Chandra NASA X rays embarked telescope, sees in near X-rays and in UV, it was observed also by the Hubble satellite  and	terrestrial facilities, like the 6.5-meter Magellan telescope from Las Campanas, Chile and the ESO telescope of Paranal Observatory (see also \cite{Clowe06}). The right hand image is in false colours, blue for X or UV rays, orange-red for matter. The Dark Matter is partially depicted by naked greenish contours, and it has been recorded, as usual by means of  the deformations of the images of the more distant galaxies. One should  notice that the blobs of Dark Matter, probably having lesser `viscosity', have overtaken the visible matter, mainly hot gas, in yellow or orange colour. Consequently, we have a succession of four different `objects' -- an invisible blob, a visible cluster of galaxies, another cluster of galaxies, and the invisible blob corresponding to this second cluster.
\cite{Clowe06}\label{fig_bullet_cluster}
}
\end{figure}
\begin{figure}[tbp]
\centering
\includegraphics[scale=0.2]{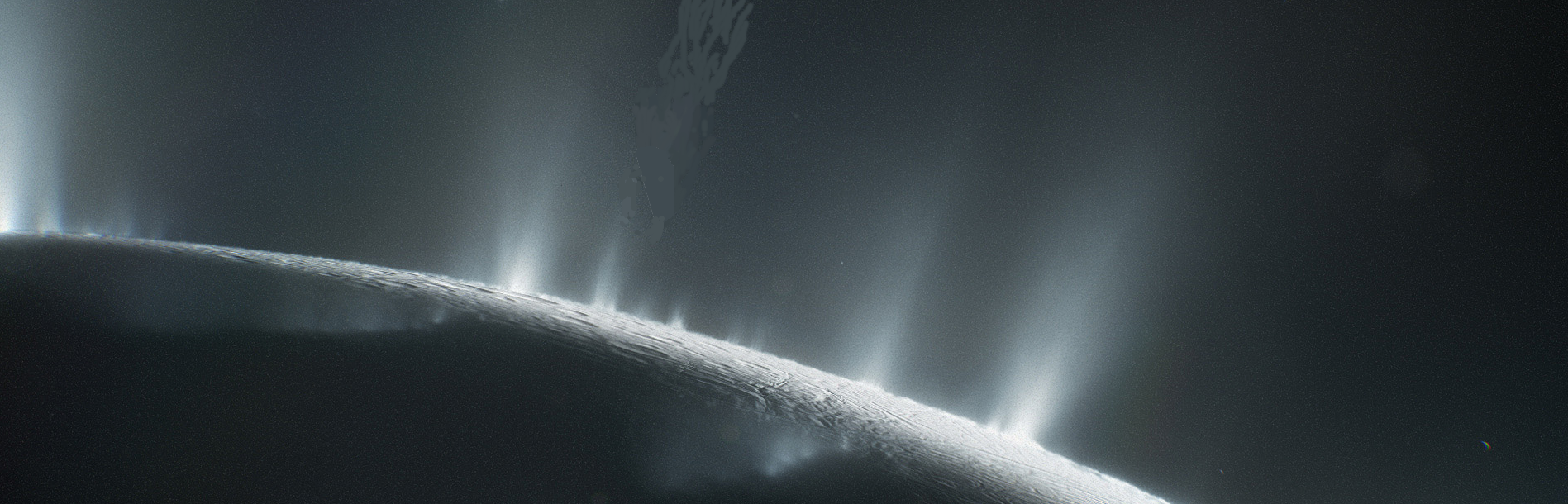}
\caption{Giant Geyser Plumes over Enceladus Southern hemisphere (courtesy NASA/JPL-Caltech). Their height is certainly  a consequence of the weakness of the gravitational field around Enceladus, but they represent  also a  direct proof  that under the ice crust the temperature (produced by tidal frictions due to the attraction of Saturn) is at least 0° Celsius, and hence compatible with Life as we know it. (Source: NASA). \label{fig_geyser_enceladus}
}
\end{figure}

\acknowledgments

The authors are thankful to the Cassini Team of the NASA Expedition to the Saturn Rings region for the multiple scientific information they have shared with us. The credits of some of the images, especially figures \ref{fig_cassini_saturn_edge_on}, \ref{fig_cassini_saturnrings}, \ref{fig_saturn_rings}, \ref{fig_bullet_cluster} and \ref{fig_geyser_enceladus} are related to them, more precisely  to Professors  Linda and Thomas Spilker from  the Jet Propulsion Laboratory, Pasadena, USA..

The authors are very thankful to Professor Guy Auberson from the Charles Coulomb Laboratory, University of Montpellier, FRANCE, and to Professor Michael Pidcock, from Oxford U.K.,  for long discussions about  these subjects. Professor Pidcock played  also a considerable role in the final form of this paper as well as in the letters exchanged with the people from NASA.  

Finally we would also like to acknowledge that we are grateful to Professor Julien Lesgourgues from Aix la Chapelle, particularly for discussions related to our first paper, \cite{sebuciulli},  concerning the points of Lagrange.


\end{document}